\documentclass[fleqn,11pt,twoside]{article}

\usepackage{amsthm,amsthm,amssymb, color, xcolor,epsfig, graphics, subfigure}
\usepackage{booktabs}
\usepackage{amsmath, graphicx, latexsym, lscape}

\makeatletter
\newcommand{\copyrightnote}[2]{{\renewcommand{\thefootnote}{}
 \footnotetext{\small\it
\begin{flushleft}
 \copyright \ #1   #2  
\end{flushleft}}}}

\newcommand{\ra}[1]{\renewcommand{\arraystretch}{#1}}
\newcommand{\Name}[1]{\begin{flushleft}
                       \LARGE \bf #1
                       \end{flushleft}\vspace{-3mm}}

\newcommand{\Author}[1]{\begin{flushleft}
                       \it #1 \end{flushleft}}

\newcommand{\Address}[1]{\begin{flushleft}
                       \it #1 \end{flushleft}}

\newcommand{\Date}[1]{\begin{flushleft}
                      \small  \it #1 \end{flushleft}}

\newcommand{\evenhead}{Author \ name}
\newcommand{\oddhead}{Article \ name}

\renewcommand{\@evenhead}{
\hspace*{-3pt}\raisebox{-15pt}[\headheight][0pt]{\vbox{\hbox to \textwidth
{\thepage \hfil \evenhead}\vskip4pt \hrule}}}
\renewcommand{\@oddhead}{
\hspace*{-3pt}\raisebox{-15pt}[\headheight][0pt]{\vbox{\hbox to \textwidth
{\oddhead \hfil \thepage}\vskip4pt\hrule}}}
\renewcommand{\@evenfoot}{}
\renewcommand{\@oddfoot}{}

\long\def\@makecaption#1#2{%
  \vskip\abovecaptionskip
  \sbox\@tempboxa{\small \textbf{#1.}\ \ #2}%
  \ifdim \wd\@tempboxa >\hsize
    {\small \textbf{#1.}\ \ #2}\par
  \else
    \global \@minipagefalse
    \hb@xt@\hsize{\hfil\box\@tempboxa\hfil}%
  \fi
  \vskip\belowcaptionskip}

\newcommand{\JNMPnumberwithin}[3][\arabic]{%
  \@ifundefined{c@#2}{\@nocounterr{#2}}{%
    \@ifundefined{c@#3}{\@nocnterr{#3}}{%
      \@addtoreset{#2}{#3}%
      \@xp\xdef\csname the#2\endcsname{%
        \@xp\@nx\csname the#3\endcsname .\@nx#1{#2}}}}%
}

\newcommand{\resetfootnoterule} {
  \renewcommand\footnoterule{%
  \kern-3\p@
  \hrule\@width.4\columnwidth
  \kern2.6\p@}
}

\renewcommand{\footnoterule}{}

\makeatother

\theoremstyle{definition}

\newtheorem{exmp}{Example}[section]

\begin{document}

\renewcommand{\evenhead}{ {\LARGE\textcolor{blue!10!black!40!green}{{\sf \ \ \ ]ocnmp[}}}\strut\hfill 
H Aratyn, J F Gomes, G V Lobo and A H Zimerman
}
\renewcommand{\oddhead}{ {\LARGE\textcolor{blue!10!black!40!green}{{\sf ]ocnmp[}}}\ \ \ \ \  
Two-fold degeneracy of a class of rational Painlev\'e V
solutions
}

\thispagestyle{empty}
\newcommand{\FistPageHead}[3]{
\begin{flushleft}
\raisebox{8mm}[0pt][0pt]
{\footnotesize \sf
\parbox{150mm}{{Open Communications in Nonlinear Mathematical Physics}\ \ \ \ {\LARGE\textcolor{blue!10!black!40!green}{]ocnmp[}}
\quad Special Issue 2, 2024\ \  pp
#2\hfill {\sc #3}}}\vspace{-13mm}
\end{flushleft}}

\FistPageHead{1}{\pageref{firstpage}--\pageref{lastpage}}{ \ \ }

\strut\hfill

\strut\hfill

\copyrightnote{The author(s). Distributed under a Creative Commons Attribution 4.0 International License}

\begin{center}

{\bf {\large Proceedings of the OCNMP-2024 Conference:\\ 

\smallskip

Bad Ems, 23-29 June 2024}}
\end{center}

\smallskip

\Name{Two-fold degeneracy of a class of rational Painlev\'e V
solutions}

\Author{H. Aratyn$^{\,1}$, J.F. Gomes$^{\,2}$, G.V. Lobo$^{\,2}$
and A.H. Zimerman$^{\,2}$}

\Address{$^{1}$ Department of Physics, 
University of Illinois at Chicago, 
845 W. Taylor St.
Chicago, Illinois 60607-7059, USA\\[2mm]
$^{2}$ Instituto de F\'{\i}sica Te\'{o}rica-UNESP,
Rua Dr Bento Teobaldo Ferraz 271, Bloco II,
01140-070 S\~{a}o Paulo, Brazil}

\Date{Received March 8, 2024; Accepted April 11, 2024}

\setcounter{equation}{0}

\begin{abstract}

\noindent 
We present a construction of a class of 
rational solutions of the Painlev\'e V Hamilton equations that 
exhibit a two-fold degeneracy, meaning that there exist two distinct
solutions that share identical parameters. 

The fundamental object of our study is the orbit of translation operators of 
the $A^{(1)}_{3}$ affine Weyl group acting on the underlying seed
solution that only allows action of some symmetry operations.  
By linking points on this orbit to rational solutions, we establish conditions for such degeneracy to occur after involving
in the construction additional B\"acklund transformations that are 
inexpressible as translation operators. 
This approach enables us to derive explicit expressions for these 
degenerate solutions. An advantage of this formalism is that it 
easily allows generalization to higher
Painlev\'e systems associated with dressing chains of even period $N>4$.

\end{abstract}

\label{firstpage}

\section{Introduction }

Painlev\'e equations are second order nonlinear
differential equations with solutions without any 
movable critical singularities in the complex plane, a
property referred to as Painlev\'e property (see e.g. \cite{gromak}). 
These solutions are generally not solvable in terms of
elementary functions however  for special values of the underlying
parameters the Painlev\'e equations possess  rational and hypergeometric-type
of solutions.

Although the discovery of
Painlev\'e equations has its origin in, mathematically motivated,
search for equations satisfying
the Painlev\'e property,  these equations and their solutions found
many practical applications and play 
an important role in several branches of mathematical physics, 
algebraic geometry, applied mathematics,fluid dynamics and 
statistical mechanics.  A list of the areas where the Painlev\'e equations
found their applications 
includes correlation functions of the
Ising model, random matrix theory, plasma physics, asymptotics of nonlinear 
partial differential equations,
quantum cohomology, conformal field theory, general relativity, nonlinear
and fiber optics, Bose-Einstein condensation 
\cite{gromak,winternitz}.
Special solutions,
such as rational solutions, turned out to play key role in these
applications and various methods were applied in their study. 

This project is dedicated to the study of rational solutions of
Painlev\'e V equation  by presenting an approach that
finds conditions for existence of degeneracy of these solutions,
derives systematically their form and also explains in a fundamental
way the origin of degenracy in the setting
of Painlev\'e V Hamiltonian formalism.  

Painlev\'e V equation
is invariant under the extended  affine Weyl group $A^{(1)}_{3}$ of
B\"acklund transformations \cite{noumibk}. A central object of our study
is a commutative subgroup of translation operators of $A^{(1)}_{3}$ and
an orbit formed by their actions on two different types of seed
solutions, one being invariant under an internal automorphism $\pi$ 
of $A^{(1)}_{3}$. 

In a recent paper \cite{AGLZ}, we have  shown how by acting with 
translation operators on a seed solution, which is invariant under automorphism 
$\pi$, one obtains Umemura
polynomials for Painlev\'e V equation 
and their relevant recurrence relations \cite{noumi-umemura}.
For the other remaining seed solution we have shown that only 
actions by selected translation operators are allowed while the remaining 
translation operators produce divergencies. 

The presence of degeneracy for a family of rational solutions 
of Painlev\'e V equation was recently pointed 
out in \cite{CD}, which also presented an explicit construction
of special function solutions in terms of the generalized Laguerre polynomials.

The novelty of our contribution is that here
we link the origin of degeneracy of rational solutions  to existence of 
divergencies resulting from actions of various translation operators
and B\"acklund transformations 
on the underlying seed solution and use it to explicitly
construct the two-fold degenerated solutions of the Painlev\'e V
Hamilton equations \eqref{hameqs} and resulting degeneracy of the Painlev\'e V
equation \eqref{finalPV} and to find 
the underlying consistency relations that dictate values of 
the parameters of degenerated solutions (see  also
a preprint \cite{AGLZ-degeneracy} for initial study of such approach).
The degeneracy of rational solutions of Painlev\'e V equation \eqref{finalPV} 
can be linked to its invariance under the simultaneous $\gamma\to -\gamma,
x \to -x$ transformation. However on the level of Hamiltonian formalism
with two canonical variables which are not both transfroming trivially
under $\gamma\to -\gamma, x \to -x$ we find that the right framework is
provided by the method that employs the translation operators and
their orbits presented in this paper. In addition this formalism lends
itself to be applied  to study of degeneracy for higher Painlev\'e systems
with the $A^{(1)}_{2k+1}, k>1$ affine Weyl symmetry group as understood on
basis of their connections with higher dressing chains of even
periodicity \cite{AGLZ}.

In section \ref{section:background}, we present the  Hamiltonian approach to Painlev\'e V 
equation and discuss the construction of rational solutions by actions
of translation operators. We describe 
solutions formed out by actions with $T_2^{-n_2}$
and $T_4^{n_4}$ translation operators, with $n_i, i=2,4$ being positive
integers, on the seed solution :
 \begin{equation}
\vert q=z,\, p=0  \rangle_{\alpha_{\mathsf{a}}} \, ,
\label{vacuum}
\end{equation}
that describes a solution of Hamilton equations  \eqref{hameqs} with values of $q,p$ being $q=z, \, p=0$ 
and an arbitrary parameter $\mathsf{a}$ equal to $\alpha_1$ and 
with zero parameters $\alpha_2$ and $\alpha_3$.
We find the recurrence relation that allows finding explicitly 
the solutions derived from \eqref{vacuum}
and obtain a close expression for their parameters 
in terms of $\mathsf{a}$ and integers $n_i, i=2,4$.

In section  \ref{section:degeneracy}, we explain a reason for existence
of degenerated solutions in the Hamiltonian formalism due to infinities associated with actions 
of some B\"acklund transformations on the seed solution \eqref{vacuum} and
use this observation to find the class of parameters that are being shared
by a pair of different in form solutions. 
We will show that degeneracy occurs for some rational solutions derived from 
\eqref{vacuum} 
for the parameter $\mathsf{a}$ that happens to be an even integer.
We propose an 
explicit construction of such solutions for a 
B\"acklund transformation $M$ 
such that infinity is generated if we are to set 
two sides of inequality 
\begin{equation}
	M  \mathbb{T}( n_2,n_4;  \mathsf{a}) 
	\ne
	\mathbb{T}( m_2,m_4;  \mathsf{b})\, ,   \; n_i,m_i \in
	\mathbb{Z}_{+}, \;i=2,4\, ,
	\label{Minequality}
\end{equation}
to be equal. 
This potential divergence is the cause of degeneracy.
In relation \eqref{Minequality}, the notation is such 
that $
\mathbb{T} ( n_2,n_4; \mathsf{a})=  T_2^{-n_2} T_4^{n_4} \vert q=z,\, p=0  
\rangle_{\alpha_{\mathsf{a}}}$ is a solution linked to the orbit of 
the seed solution \eqref{vacuum} under actions of $T_2$ and $T_4$
operators. 
To be responsible for degeneracy the  
B\"acklund transformation $M$ must be such that it satisfies two
conditions. First that it will cause the
divergence, 
as described in equation \eqref{Mequationc}, and secondly 
that the equation 
 \begin{equation}
M  \left(\alpha_{n; \mathsf{a}}\right) = \alpha_{m ; \mathsf{b}}\, ,
\label{Mequality}
\end{equation}
with the symbol $\alpha_{n; \mathsf{a}}=(\mathsf{a} +2 n_2,\, -2 n_2,\,
 -2 n_4,\,  2-\mathsf{a} +2 n_4)$ (see relation  \eqref{orbit}),
will have a solution for some values of the
parameters $n_i, m_i, i=2,4$ and $\mathsf{a}, \mathsf{b}$
ensuring that both sides of inequality \eqref{Minequality}
will share the same parameter.
These two conditions are shown to be satisfied
for $M$ being one of the  B\"acklund transformations 
$M_{12}=s_1 s_2$, $M_{34}=s_3 s_4$, $M_1=\pi s_1 $, $M_4=\pi^{-1}  s_4$
and we call the corresponding set of degenerated solutions an
$M_i$-sequence.
One of the main points of this paper is 
that all these four sequences 
are equivalent. 
Specifically, the sequences  $M_1, M_{12}$ and $M_4$
are mapped  into each other by B\"acklund transformations,
while $M_{3,4}$ happens to be equivalent to $M_1$ after a
simple re-definitions of underlying  parameters
as discussed in subsections \ref{subsection:M4} - \ref{subsection:M12}.
The equivalence of these sequences 
is a new result not contained in unpublished reference \cite{AGLZ-degeneracy}.

The final section \ref{section:discussion}, offers conclusions and 
discussion of the results. This section 
reviews the results shown in Examples \ref{exmp:n2=2n4=2}, \ref{exmp:n2=3n4=1}
and \ref{exmp:n2=1n4=3} to obtain an unifying discussion for special 
values of parameters labeling the degenerated solutions of the Hamilton
Painlev\'e V equations.
We find that the condition for a solution
constructed in section \ref{section:background} to be equal to
one of the degenerated solutions is that the underlying parameter 
$\mathsf{a}$ of the seed solution is an even integer.
We also remark that the fact that the discussion of degeneracy of
Painlev\'e systems is here placed firmly in the setting of  the extended  affine Weyl group 
$A^{(1)}_{N-1}, N=4$ lends itself naturally to being  generalized to 
Painlev\'e systems associated with 
higher dressing chains of even period $N>4$, where more richer
degeneracy structure is expected to appear.

\section{Background}
\label{section:background}
We will mainly be working with the  Hamiltonian approach to Painlev\'e V equation
with the Hamilton:
\begin{equation}
H= -q\,(q-z)\,p\, (p-z)+(1-\alpha_1-\alpha_3)\, p q +\alpha_1 z
p-\alpha_2 z q \, ,
\label{pqHam}
\end{equation}
where $\alpha_i, i=1,2,3$ are three constant parameters and
$q, p$ are two canonical variables that satisfy Hamilton equations: 
$z q_z = {d H}/{d p}$ , $z p_z =- {d H}/{d q}$:
\begin{equation}\begin{split}
z q_z &= -q(q-z)(2p-z)+(1-\alpha_1-\alpha_3)  q +\alpha_1 z \, ,\\
z p_z &= p (p-z)(2q-z) - (1-\alpha_1-\alpha_3) p +\alpha_2 z \, ,
\label{hameqs}
\end{split}
\end{equation}
from which one derives Painlev\'e V equation 
\begin{equation} 
y_{x x}
= -\frac{y_x}{x}+\left( \frac{1}{2y}+ \frac{1}{y-1}\right)
y_{x}^2 + \frac{(y-1)^2}{x^2} \left({ \alpha} y + 
\frac{{ \beta}}{y} \right) + \frac{{ \gamma}}{ x} y + { \delta} 
\frac{y (y+1)}{y-1}\, ,
\label{finalPV}
\end{equation}
by eliminating one of the
canonical variables and defining $y=(q/z) (q/z-1)^{-1}$, 
as well as redefining the variable $z \to x$ with 
$x = \epsilon z^2/2$.
The coefficients $\alpha,\beta,\gamma$ 
of the  Painlev\'e V equation are given by:
\begin{equation} 
{ \alpha}= \frac18 \alpha_3^2,\;\; { \beta}=- \frac18 \alpha_1^2,\;\;
{ \gamma}=  \frac{\alpha_2-\alpha_4}{2 \epsilon },\;\;
{ \delta}=   -\frac12 \frac{1}{\epsilon^2}\, ,
\label{paramPVbar}
\end{equation}
in terms of 
components $\alpha_i=(\alpha_1,\alpha_2, \alpha_3,\alpha_4)$ 
with $\alpha_4=2-\sum_{i=1}^3 \alpha_i$.

For ${ \delta}$ to take a conventional value of $-\frac12$ we need 
$ \epsilon^2 = 1$.

The Hamilton equations are directly connected to symmetric  Painlev\'e
V equations:
\[
z \frac{d f_i}{d z}=
f_i f_{i+2} \left(f_{i+1}-f_{i-1}\right) +
\left(1-\alpha_{i+2}\right) f_i + \alpha_i f_i \, , \;\;f_{i+4}=f_i
, \;\;i=1,2,3,4\, ,
\]
via relations $f_1=q,\, f_2=p,\, f_3=z-q,\,f_4=z-p$. Since our
formalism will be shown to describe degeneracy of Painlev\'e
V Hamilton equations \eqref{hameqs} it will also automatically
provide such description for the symmetric  Painlev\'e
V equations as well as equation \eqref{finalPV}.

The Hamilton equations are invariant
under B\"acklund transformations, $\pi, s_i,i=1,{\ldots}, 4$
that satisfy the $A^{(1)}_{3}$ extended affine Weyl group relations:
\begin{xalignat}{2}
   s_i^2=1, &\qquad \quad s_i s_j =s_j s_i~(j \ne i,i \pm 1), &
   s_i s_j s_i =s_j s_i s_j~(j=i \pm 1),\nonumber \\
  \pi^4=1, & \qquad \quad \pi s_j =s_{j+1}\pi,\;\; \;\; s_{i+4}=s_i.&
\label{fun.rel}
\end{xalignat}
An explicit form of these transformations
 on canonical variables $p$ and $q$ is shown
in Table \ref{backsymham},
\begin{table}[h]
	\centering
	\ra{1.8}
	\begin{tabular}{ccccccc}
		\toprule
		& $q$ & $p$ & $\alpha_1$ & $\alpha_2$ & $\alpha_3$ & $\alpha_4$ \\
		\midrule
		$s_1$ & $q$ & $p + \frac{\alpha_1}{q}$ & $-\alpha_1$ & $\alpha_1 + \alpha_2$ & $\alpha_3$ & $\alpha_1 + \alpha_4$ \\
		$s_2$ & $q- \frac{\alpha_2}{p}$ & $p$ & $\alpha_1 + \alpha_2$ & $-\alpha_2$ & $\alpha_2 + \alpha_3$ & $\alpha_4$ \\
		$s_3$ & $q$ & $p -\frac{\alpha_3}{z-q}$ & $\alpha_1$ & $\alpha_2+\alpha_3$ & $- \alpha_3$ & $\alpha_3 + \alpha_4$ \\
		$s_4$ & $q + \frac{\alpha_4}{z-p}$ & $p$ & $\alpha_1 + \alpha_4$ & $\alpha_2$ & $\alpha_3 + \alpha_4$ & $-\alpha_4$ \\
		$\pi$ & $z-p$ & $q$ & $\alpha_4$ & $\alpha_1$ & $\alpha_2$ & $\alpha_3$ \\
		\bottomrule
	\end{tabular}
	\caption{$A^{(1)}_{3}$ B\"acklund transformations}
	\label{backsymham}
\end{table}
 Imposing the periodicity
condition $ \alpha_{i+4}=\alpha_{i}$ we can compactly describe 
the action of the B\"acklund transformations 
on the  constant parameters $\alpha_i$ 
from equations \eqref{hameqs} as :
\begin{equation}
s_i (\alpha_i) = -\alpha_i, \quad
s_i (\alpha_{i \pm 1}) = 
 \alpha_i+\alpha_{i \pm 1}, \quad
s_i (\alpha_{i +2}) = 
 \alpha_{i +2}, \quad i=1,2,3,4 \, .
\label{sialpha}
\end{equation}
Furthermore  the automorphism $\pi$ acts according to
\begin{equation}
\pi (\alpha_i)  = \alpha_{i-1}  \, .
\label{pialpha}
\end{equation}
 
Within the $A^{(1)}_{3}$
extended affine Weyl group one defines an abelian 
subgroup of translation operators defined as 
$T_i=r_{i+3} r_{i+2} r_{i+1} r_i, i=1,2,3,4$, 
where $r_i=r_{4+i}=s_i$ for $i=1,2,3$ and $r_4 = \pi$.
The translation operators commute among themselves, 
$ T_i T_j =T_j T_i$, and as follows from relations
\eqref{sialpha} and \eqref{pialpha} 
generate the following translations 
when acting on the $\alpha_i$ parameters:
\[ T_i (\alpha_i) = \alpha_i+2,\;  T_i (\alpha_{i-1})=\alpha_{i-1}-2,\;
 T_i (\alpha_j) =  \alpha_j, \; j=i+1, j=i+2\,.\]
The translation
operators satisfy the following commutation relations 
\begin{equation} 
s_i T_i s_i =T_{i+1},\;\,  s_i T_j s_i=T_j
, \,j\ne i, i+1, \;\;  \pi\; T_i =T_{i+1}\, \pi \,,
\label{lemmasi}
\end{equation}
with the B\"acklund transformations $s_i, \, i=1,2,3,4$ and an automorphism
$\pi$ and the usual periodicity condition $T_{i+4}=T_i$ being imposed.

The reference \cite{AGLZ} described construction of  rational solutions
of Painlev\'e V equation  out of actions of translation  operators
on seed solutions that first appeared in \cite{watanabe}. 
Crucial for this construction is that rational solutions fall into two classes 
depending on which of the two types
of seed solutions they have been derived
from by actions of translation operators.
These two classes of seed solutions are: \begin{enumerate}
\item $q=z/2,\,p=z/2$, with the
parameter $\alpha=(\mathsf{a},1-\mathsf{a},\mathsf{a},1-\mathsf{a})$\,,
\item   $q=z,\, p=0$, with the
parameter $\alpha_{\mathsf{a}}=(\mathsf{a},0,0,2-\mathsf{a})$ denoted
here by
$\vert q=z,\, p=0  \rangle_{\alpha_{\mathsf{a}}}$.
\end{enumerate}
They both  solve
the Hamilton equations \eqref{hameqs}  
for an arbitrary variable $\mathsf{a}$.
As shown in  \cite{AGLZ},
the first class of seed solutions gives rise to
Umemura polynomials and the second to special functions. It was also
shown there that the solutions constructed with this procedure 
satisfy all sufficient and necessary conditions  for
the parameters of rational solutions of Painlev\'e V equation first derived in
\cite{kitaev}.
The action of the B\"acklund transformation $s_i$ on the seed solution
\eqref{vacuum} is :
\[ \big\vert q=z,\, p=0  \big\rangle_{\alpha_{\mathsf{a}}}
\stackrel{s_i}{\longrightarrow}
\big\vert s_i (q=z),\, s_i(p=0)  \big\rangle_{s_i
(\alpha_{\mathsf{a}})} \, ,
\]
and similarly for all the other B\"acklund transformations.

Acting repeatedly with the $\pi$ automorphism on the seed solution \eqref{vacuum} 
produces three other variants of such
 solution. They all serve as seed solutions in analogous way to the 
 solution \eqref{vacuum}.
 Here we will limit our discussion  only to the seed solution 
\eqref{vacuum} and solutions generated from it
as
the other solutions and the corresponding structure of degeneracy
follow from the same formalism under 
appropriate actions of $\pi$.

The B\"acklund transformations $s_2,s_3$ generate infinity when applied
on the solution \eqref{vacuum} and accordingly only
actions by some powers of $T_1,T_2, T_4$ are well defined on a seed
solution $\vert q=z,\, p=0  \rangle_{\alpha_{\mathsf{a}}}$.
The allowed operations 
are as follows \cite{AGLZ}:
\[ T_1^{n_1} T_2^{-n_2} T_4^{n_4} \vert q=z,\, p=0  \rangle_{\alpha_{\mathsf{a}}}, \; n_1
\in \mathbb{Z},  \; n_2,n_4 \in \mathbb{Z}_{+} \,.\]
This operation is to be understood as producing
new solutions $q$ and $ p$ of the Hamilton equations 
equal to
$ T_1^{n_1} T_2^{-n_2} T_4^{n_4} (q=z)$ and 
$ T_1^{n_1} T_2^{-n_2} T_4^{n_4} (p=0)$
and with a new
parameter:
\begin{equation}  T_1^{n_1} T_2^{-n_2} T_4^{n_4} (\alpha_{\mathsf{a}} )
=(\mathsf{a} +2 n_1+2 n_2,\, -2 n_2,\,
 -2 n_4,\,  2-\mathsf{a} +2 n_4-2 n_1)
\,.
\label{alphaT1T2T4}
\end{equation} 
Evidently, the action of $T_1^{n_1} $ only amounts to shifting a 
parameter   $\mathsf{a}$ and as shown in \cite{AGLZ} leaves  the 
configuration $q=z, p=0$ unchanged.  Thus : 
\begin{equation}
T_1^{n_1} \vert q=z,\, p=0  
\rangle_{\alpha_{\mathsf{a}}}= \vert q=z,\, p=0  
\rangle_{\alpha_{\mathsf{a}+2 n_1}}.
\label{T1shift}
\end{equation}
We can therefore, 
largely, ignore $T_1$ and restrict our discussion to the solutions
of Painlev\'e V equation of the form :
\begin{equation} \begin{split}
\mathbb{T} ( n_2,n_4; \mathsf{a})&=  T_2^{-n_2} T_4^{n_4} \vert q=z,\, p=0  
\rangle_{\alpha_{\mathsf{a}}},  \; n_2,n_4 \in \mathbb{Z}_{+} \, ,\\
\alpha_{n; \mathsf{a}}&=  T_2^{-n_2} T_4^{n_4} (\alpha_{\mathsf{a}} )
=(\mathsf{a} +2 n_2,\, -2 n_2,\,
 -2 n_4,\,  2-\mathsf{a} +2 n_4)\, ,
\end{split}
\label{orbit}
\end{equation}
where we listed both the solution generated by translation operators
and its corresponding  parameter  $ \alpha_{n, \mathsf{a}}$.
$\mathbb{Z}_{+}$ contains positive integers and zero.

To describe solutions $\mathbb{T} ( n_2,n_4; \mathsf{a})$
we will first set $n_4=0$ and recall expressions for an action 
by $T_2^{-n}$ \cite{AGLZ}:
\begin{equation}
\begin{split}
T_2^{-1}&: \big\vert q=z,\, p=0  \rangle_{\alpha_{\mathsf{a}}} \to 
\vert q=z, p= \frac{2z}{\mathsf{a} -z^2}
\big\rangle_{(2+\mathsf{a},-2,0,2-\mathsf{a})}\\
T_2^{-n}&: \big\vert q=z,\, p=0  \rangle_{\alpha_{\mathsf{a}}} 
\to \vert q_n=z, p_n=\frac{2n z R_{n-1} (x,\mathsf{a})}{R_{n}
(x,\mathsf{a}) }\big\rangle_{(\mathsf{a}+2n,-2n,0,2-\mathsf{a})}\, ,
\label{T2n}
\end{split}\end{equation}
where $x=-z^2/2$ and $R_{n} (x, \mathsf{a})$ are Kummer polynomials
that satisfy the recurrence relations:
\begin{align}
2 k R_{k-1}(x,\mathsf{a})&= R_k(x,\mathsf{a})-R_k(x,\mathsf{a}-2)=
 \frac{d R_k (x,\mathsf{a}) }{d x} \, ,
\label{Rrecur1}\\
R_{k+1} (x,\mathsf{a})&= 2 x R_{k} (x,\mathsf{a})+\mathsf{a} R_{k}
(x,\mathsf{a}+2) \, ,
\label{Rrecur2}
\end{align}
for $k=0,1,2,{\ldots} $ with $R_0 (x,\mathsf{a})=1$ (see e.g.
\cite{AGLZ-degeneracy,AGLZ-orbit}).

The result for $T_2^{-n_2} \vert q=z,\, p=0  
\rangle_{\alpha_{\mathsf{a}}}$ is obtained
by inserting $n=n_2$ into equation \eqref{T2n}.

The further  action  with $T_4^{n_4}$ utilizes expression
\begin{equation}
\begin{split}
T_4(q)&= z-p-(\alpha_1+\alpha_4)/(q+\alpha_4/(z-p))\, , \\
T_4 (p) &=
q+\alpha_4/(z-p)-(\alpha_1+\alpha_2+\alpha_4)/(p+(\alpha_1+\alpha_4)/(q+\alpha_4/(z-p)))
\, ,
\label{T4pq}
\end{split}
\end{equation}
describing action of the translation operator $T_4$  on a solution
$q,p$ of the Hamilton equations \eqref{hameqs} 
with $\alpha_i=(\alpha_1,\alpha_2,\alpha_3,\alpha_4)$.
The recurrence relations obtained from expression \eqref{T4pq} are:
\begin{equation}
\begin{split}
q^{(k)}&=T_4^{k} (q_0) = z- p^{(k-1)} - \frac{2(k+n_2)}{v_k}
=z - u_k\, .
\\
p^{(k)}&=T_4^{k} (p_0) = v_k - \frac{2k}{u_k}\, .\\
 \alpha^{(k)}&=(\mathsf{a} +2 n_2,\, -2 n_2,\,
 -2 k,\,  2-\mathsf{a} +2 k), \;\;\,k=1,2,{\ldots} , n_4\, ,
\label{T4k}
\end{split}
\end{equation}
where \[v_k =q^{(k-1)}+\frac{2k-\mathsf{a}}{z - p^{(k-1)}}, \;\; \;
u_k =p^{(k-1)}+\frac{2(k+n_2)}{v_{k}},\]
and $q_0=z$ and $p_0=\frac{2n z R_{n_2-1} (x,\mathsf{a})}{R_{n_2}
(x,\mathsf{a}) }$. Setting $k=n_4$ into $\alpha^{(k)}$ we recover 
$\alpha_{n; \mathsf{a}}$ from expression 
\eqref{orbit}. The closed expressions for $q^{(k)},p^{(k)}$ will be
described in the future publication \cite{AGLZ-orbit}.

In the next section we will derive the parameters of degenerated
solutions (see e.g. \eqref{alphapi1}) and compare with the above value
of the parameter $\alpha_{n; \mathsf{a}}$ on the orbit of 
$T_2^{-n_2} T_4^{n_4} $. In section \ref{section:discussion} we will
find that for any $\mathsf{a}$ that is 
an even integer the parameter $\alpha_{n; \mathsf{a}}$ can be cast
in a form of a parameter of degenerated pair of
solutions.

\section{Degeneracy}
\label{section:degeneracy}
The above construction of solutions in section \ref{section:background}
did not take into account existence of any
other B\"acklund transformations than translation operators. The B\"acklund transformations
that are not expressible in terms  of translation operators will
play a role in what follows.
Our construction associates the (two-fold) degeneracy 
to inequality \eqref{Minequality} with two sides that are  
two different (finite) solutions  of Painlev\'e V Hamilton 
equations that share a common   Painlev\'e V parameter \eqref{Mequality}.

In relations \eqref{Minequality}  and \eqref{Mequality}
the symbol $M$ denotes a B\"acklund transformation, which is not expressible
in terms of translation operators only and
such that $ T_2^{m_2} T_4^{-m_4} M  \mathbb{T}( n_2,n_4; \mathsf{a})$ 
is ill-defined as we will see below.
For that reason the two solutions listed in \eqref{Minequality}
can not be equal. 
We will refer
to degenerated solutions of relations \eqref{Minequality} and 
\eqref{Mequality} as $M$-sequence.

To determine general  conditions for degeneracy let us equate for the moment
expressions on the left and the right sides of the inequality \eqref{Minequality} 
with each other and multiply both sides
with $T_2^{m_2} T_4^{-m_4}$ to get:
\[
 \vert q=z,\, p=0  
\rangle_{\alpha_{\mathsf{b}}}
=  T_2^{m_2} T_4^{-m_4} M T_2^{-n_2} T_4^{n_4}\,
 \vert q=z,\, p=0  
\rangle_{\alpha_{\mathsf{a}}}= 
M \,T_3^{c_3} \,T_2^{c_2}\, T_4^{c_4}  \vert q=z,\, p=0  
\rangle_{\alpha_{\mathsf{a}}}
\]
obtained after commuting  $T_2^{m_2} T_4^{-m_4}$ around $M$ and ignoring
potential presence of $T_1$ on the right hand side since it only amounts
to shifting of $\mathsf{a}$.
The conditions for degeneracy in this setting are 
\begin{equation}
	c_3 \ne 0 , \;\, \text{or}\;\, c_2 >0\, , \;\, \text{or}\;\, c_4 <0 
	\,,
	\label{Mequationc}\end{equation}
since they correspond to presence of operators that will cause divergence
when acting on  $\vert q=z,\, p=0  
\rangle_{\alpha_{\mathsf{a}}}$.
We next explore several candidates for $M$ to see if they satisfy the conditions
\eqref{Mequality} and \eqref{Mequationc}.

We can easily discard $M=s_2,M=s_3$ as they do not satisfy  the condition 
\eqref{Mequality}, as it would require $m_2=-n_2$ for $s_2$ and
$m_4=-n_4$ for $s_3$. Further,  one finds
that  $M=s_1,M=s_4$ do not produce infinities 
and accordingly fail to satisfy the conditions of relation
\eqref{Mequationc}.

Moving on to the quadratic expressions of the type $s_i s_j$
we find that when $j \ne i+1$ (e.g. $s_1 s_3$ or $s_2 s_4$)
then both expressions do not  satisfy the condition
\eqref{Mequality}. The remaining cases are of the type $s_i s_{i+1}$
since $s_i s_{i-1}$ can be moved from the left to the right hand side
of relation \eqref{Minequality} to become
$s_i s_{i+1}$. Inspection of $s_1 s_2,s_2s_3,s_3 s_4, s_4 s_1$ shows
that only 
\begin{enumerate}
\item $M_{12}=s_1 s_2$,
\item 
$M_{34}=s_3 s_4$,
\end{enumerate}
satisfy the condition 
\eqref{Mequality} and the condition \eqref{Mequationc} for some values of
$m_i,\, i=2,4$.
These conditions are also satisfied by 
\begin{enumerate}
\item[3.] $M_1=\pi s_1 $,
\item[4.] $M_4=\pi^{-1}  s_4$,
\end{enumerate}
that are effectively equivalent to the cases of
$M=\pi , \pi^{-1}$ \cite{AGLZ-degeneracy}. It is also easy to see that
$M_1$ and $M_4$ are not invertible in the context of relation \eqref{Minequality}
since $M_i^{-1}, i=1,4$ acting on $\mathbb{T}( m_2,m_4;  \mathsf{b})$
will cause a divergence. Thus if an
equality between two solutions shown in \eqref{Minequality} held 
for $M_1$ or $M_4$ then an
attempt to invert $M_1$ or $M_4$ would have produced an infinity. 

It suffices to consider operators $M$ that consist of a single $s_i$
multiplied by $\pi$ or a product of two $s_i$'s
due to the following identities :
\begin{equation}
\begin{split}
s_i s_{i+1} &= \pi s_{i+2}  T_{i+2}^{-1}=\pi T_{i+3}^{-1} s_{i+2}
,\quad \; i=1,2,3,4\, ,\\
s_{i+1} s_i  &=  \pi^{-1}   s_{i-1}T_{i}=  \pi^{-1} T_{i-1}  s_{i-1}
,\quad \; i=1,2,3,4 \, ,
\end{split}
\label{nextsi}
\end{equation}
for products of neighboring $s_i$ that reduce them to one single
$s_i$ multiplied by a shift operator and an automorphism $\pi$.
Accordingly, in principle, the higher products of $s_i$ can be reduced 
to the lower
number of $s_i$ transformations \cite{AGLZ-degeneracy}.

We will now examine if there exists equivalence between
the four cases with degeneracy represented by 
$M_1,M_4, M_{12}, M_{34}$.
We choose as 
a starting point 
the relation \eqref{Mequality} with $M=M_1=\pi s_1 $ and
accordingly with the parameter :
\begin{equation} 
\pi s_1 \left(\alpha_{n;\mathsf{a}}\right)= \alpha_{m;\mathsf{b}}
= 2(1+n_2+n_4,-m_2,m_2-n_2,-n_4) \, ,
\label{alphapi1}
\end{equation}
shared between the two solutions appearing in the inequality:
\begin{equation}  \pi  s_1 \mathbb{T}\left( n_2,n_4; \mathsf{a}\right)
\ne  \mathbb{T}\left( m_2,m_4; \mathsf{b} \right)\,.
\label{Xone}
\end{equation}
Expression \eqref{alphapi1} holds when the following consistency conditions 
are satisfied :
\begin{align} 
m_4&=n_2-m_2 \ge 0, \;  n_2 \ge m_2 \ge 0, \, n_2 , m_2, n_4 
\in \mathbb{Z}_{+}\, , \label{Xoneca}\\
\mathsf{a}&=2(m_2- n_2)=-2 m_4, \; \mathsf{b}=2+2 n_4+2m_4=2+2 n_4-
\mathsf{a} \,.
\label{Xonec} 
\end{align}

\begin{exmp}
\label{exmp:n2=2n4=2}
We consider the case of  
\begin{equation} n_2=n_4=2, \;  m_2=1 \; \to m_4=n_2-m_2=1,
\, \alpha_i=2 (5,-1,-1,-2)\, ,
\label{itemA}
\end{equation}
where we used relation \eqref{alphapi1} to calculate $\alpha_i$ and
the consistency condition \eqref{Xoneca}.
For the corresponding coefficients of the Painlev\'e V equation 
we find from relation \eqref{paramPVbar} for $\epsilon=1$:
\begin{equation}
\alpha = \frac12,\; \; \beta= - \frac{25}{2},\;\; 
\gamma= 1 \, .
\label{paramPVbar1}
\end{equation}
According to rules of the $M_1$-sequence 
we have two degenerated solutions corresponding to the parameters given in
equation \eqref{itemA}: 
\begin{equation}
\begin{split}
\pi s_1 \mathbb{T} (n_2=2,n_4=2; \mathsf{a}=-2) &
=\pi s_1 T_2^{-2} T_4^{2}  
\big\vert q=z,\, p=0  \rangle_{\alpha_{\mathsf{a}=-2}} 
\, ,\\
\mathbb{T} (m_2=1,m_4=1; \mathsf{b}=8) &= 
T_2^{-1} T_4^{1}
\big\vert q=z,\, p=0  \rangle_{\alpha_{\mathsf{b}=8}} 
\, .	
\end{split}\label{Xn1Ym1}
\end{equation}
with $\mathsf{a}$ and $\mathsf{b}$ determined from relation \eqref{Xonec}.

We first calculate $\mathbb{T} (m_2=1,m_4=1; \mathsf{b}=8) $
from expression \eqref{Xn1Ym1} using the first of relations \eqref{T2n}
with the parameter $\mathsf{b}$ followed by 
action with $T_4$ according to \eqref{T4pq}
to get
\begin{equation}
\begin{split}
q&=z\frac{(-\mathsf{b}+z^2+2)(z^4-2 z^2 \mathsf{b}+\mathsf{b}^2+2\mathsf{b})}{(-\mathsf{b}+z^2)(-2z^2
\mathsf{b}+z^4+4z^2-2\mathsf{b}+\mathsf{b}^2)}\\
p&=-2z \frac{(-2 z^2 \mathsf{b}+z^4+4 z^2-2\mathsf{b}+\mathsf{b}^2)}{(-\mathsf{b}+z^2+2)(-2z^2
\mathsf{b}+z^4-2\mathsf{b}+\mathsf{b}^2)}
\label{ppqq}
\end{split}
\end{equation}
which for $\mathsf{b}=8$ yields
\begin{equation}
q= \frac{z (z^2-6) (z^4-16 z^2+80)}{(z^2-8)(z^4-12 z^2+48)}, \;\;
p =  \frac{-2 z(z^4-12 z^2+48)}{ (z^2-6)(z-2)(z+2)(z^2-12)} \, ,
\label{solutions11}
\end{equation}
with $\alpha_i= (10, -2,-2,-4)=2 (5,-1,-1,2)$.
To obtain a solution $y(x)$ of the Painlev\'e V equation we transform 
$q \to y=(q/z)(q/z-1)^{-1}$  and substitute $z$ by $x=-z^2/2$ 
with the result:
\begin{equation} 
y(x)= 
+\frac{(x+3)(x^2+8x+20)}{(x+2)(x+6)}\, ,
\label{w111}
\end{equation}
which agrees with the expression of the Painlev\'e V solution 
$w_{1,1}(x;1)$ obtained in Example 4.11 of \cite{CD}.

Next we calculate $\pi s_1 \mathbb{T} (n_2=2,n_4=2; \mathsf{a}=-2)$ 
from relation \eqref{Xn1Ym1} acting first
with $T_2^{-2}$ on  $q=z,\; p=0$
that according to equation \eqref{T2n}
for $n=2$ yields:
\begin{equation}
T_2^{-2}: q=z, p=0 \to q=z, p=\frac{4 z (\mathsf{a}-z^2)}{z^4 -2 \mathsf{a}
z^2+\mathsf{a}(\mathsf{a}+2) },  \, 
(4+\mathsf{a},-4,0,2-\mathsf{a})\, ,
\label{T2m2}
\end{equation}
Applying $T_4^2$, using expression \eqref{T4pq}, on the configuration in
equation \eqref{T2m2} we get a complicated  solution to Painlev\'e
equation for $\alpha_i=(4+\mathsf{a},-4,-4,6-\mathsf{a})$.
Inserting $\mathsf{a}=-2$ simplifies $\alpha_i$ to $(2,-4,-4,8)$
and the expressions for $q,p$ simplify  to:
\begin{equation}
\begin{split}
q&=z\,\frac{(z^4+12\,z^2+48)\,(z^8+16\,z^6+96\,z^4+192\,z^2+192)}
{(z^8+24\,z^6+216\,z^4+768\,z^2+1152)\,(8\,z^2+24+z^4)}\, ,\\
p&=-4\,\frac{(z^6+6\,z^4+24\,z^2+48)\,(z^8+24\,z^6+216\,z^4+768\,z^2+1152)}
{z\,(z^6+12\,z^4+72\,z^2+192)\,(z^8+16\,z^6+96\,z^4+192\,z^2+192)} \, ,
\end{split}
\label{solsT2T4}
\end{equation}
Applying then $\pi s_1$ that transforms : $(2,-4,-4,8)\to (10,-2,-2,-4)$ we are
being taken from solution \eqref{solsT2T4} to: 
\begin{equation}
\begin{split}
 q&=
z\,\frac{(8\,z^2+24+z^4)\,(z^6+18\,z^4+144\,z^2+480)}{(z^4+12\,z^2+48)\,(z^6+12\,z^4+72\,z^2+192)}
\, , \\
p&=z\,\frac{(z^4+12\,z^2+48)\,(z^8+16\,z^6+96\,z^4+192\,z^2+192)}{(z^8+24\,z^6+216\,z^4+768\,z^2+1152)\,(8\,z^2+24+z^4)}
\, ,
\label{solutions2}
\end{split}
\end{equation}
which, as it was the case with expressions \eqref{solutions11},
solves the Painlev\'e V Hamilton equation with $\alpha_i= (10, -2,-2,-4)$.

The corresponding    solution $y(x)=(q/z)(q/z-1)^{-1}$ of the Painlev\'e V equation
for coefficients \eqref{paramPVbar1} reads
  \begin{equation}
y = \frac{(x^2-4 x+6) (-x^3+9 x^2-36 x+60)}{x^4-12 x^3+54 x^2-96 x+72}
\, ,
\label{hatw11m1}
\end{equation}
that agrees with expression for ${\hat w}_{1,2}(x;-1)$ of Example 4.11
of reference \cite{CD}.
\end{exmp}

\begin{exmp}
\label{exmp:n2=3n4=1}
Next we consider the case of  
\begin{equation}n_2=3, \, n_4=1, \; m_2=2 \; \to m_4=n_2-m_2=1,
 \, \alpha_i=2 (5,-2,-1,-1)\, ,
\label{itemB}
\end{equation}
For the corresponding coefficients of the Painlev\'e V equation 
we find from relation \eqref{paramPVbar} for $\epsilon=1$:
\begin{equation}
\alpha = \frac12,\; \; \beta= - \frac{25}{2},\;\; 
\gamma= -1 \, .
\label{paramPVbar2}
\end{equation}
We notice that the above coefficients differ from the ones 
in equation \eqref{paramPVbar1} of Example \ref{exmp:n2=2n4=2}
only by the sign of $\gamma$, which will be of importance below.

Again, according to rules of the $M_1$-sequence 
we have two degenerated solutions corresponding to the parameters given in
equation \eqref{itemB}: 
\begin{equation}
\begin{split}
\pi s_1 \mathbb{T} (n_2=3,n_4=1; \mathsf{a}=-2) &
=\pi s_1 T_2^{-3} T_4^{1} 
\big\vert q=z,\, p=0  \rangle_{\alpha_{\mathsf{a}=-2}} 
 \, ,\\
\mathbb{T} (m_2=2,m_4=1; \mathsf{b}=6) &= 
T_2^{-2} T_4^{1}
\big\vert q=z,\, p=0  \rangle_{\alpha_{\mathsf{b}=6}} 
\, .	
\end{split}\label{BXn1Ym1}
\end{equation}
with 
$\mathsf{a}=-2 m_4=-2, \mathsf{b}=2+2 n_4+2m_4=6$.

We  first use expression \eqref{T2n} that 
gives for $n=3$ :
\begin{equation}
T_2^{-3}: q=z, p=0 \to q=z, p=\frac{6 z R_2 (x, \mathsf{a})}{R_3 (x,\mathsf{a}) },  \, 
\alpha_i = (6+\mathsf{a},-6,0,2-\mathsf{a})\, ,
\label{T2m3}
\end{equation}
where for $x=-z^2/2$:
\begin{equation}
R_1(x,\mathsf{a})= \mathsf{a}+2x, \; R_2(x,\mathsf{a})=-4 x +(2+\mathsf{a}+2x)
( \mathsf{a}+2x)
\label{R2xa}
\end{equation}
and 
\[
R_3(x,\mathsf{a})= (2x)^3+ 3 (2 x)^2 \mathsf{a} 
+ 3 (2 x) \mathsf{a} (\mathsf{a}+2)+
\mathsf{a} (\mathsf{a}+2)(\mathsf{a}+4)
\] 
as follows from the recurrence relation \eqref{Rrecur2}.
Using the transformation rule  \eqref{T4pq} 
and applying $\pi s_1$ and setting $\mathsf{a}=-2$ so 
that $\alpha_i=(10,-4,-2,-2)$ we obtain for 
the first of equations \eqref{BXn1Ym1}
\[
\begin{split}
\pi s_1 \mathbb{T} (n_2=3,n_4=1; \mathsf{a}=-2) =
(q&= \frac{z (z^6+22 z^4+176 z^2+480)}{(z^2+8)(z^4+48+12 z^2)},\\
p&= \frac{z (z^2+8)(z^4+12 z^2+24)}{(6+z^2)(z^2+4)(z^2+12)})\, ,
\end{split}
\]
which gives for  $y=(q/z)/(q/z-1)$:
\begin{equation}
y = \frac{(-x+3) (x^2-8x+20)}{(-x+6)(-x+2)}
\label{y1itemB}
\end{equation}
Note that going from Example \ref{exmp:n2=2n4=2}  to Example
\ref{exmp:n2=3n4=1}
($  \alpha_i=2 (5, -1,-1,-2) \to
 \alpha_i= 2 (5, -2,-1,-1)$) only amounts to flipping sign of $\gamma$:
 $\gamma \to -\gamma$ in the Painlev\'e V equation.
However the transformation $\gamma \to -\gamma$ amounts to
$x \to -x$. Thus we go  
from the solution \eqref{w111} of Painlev\'e V equation
to the solution \eqref{y1itemB}
only  by  flipping the sign of $x$ as it is easily verified by inspection.

Using \eqref{T2m3} and the transformation rule  \eqref{T4pq}  
we get
\[
\begin{split}
\mathbb{T} (m_2=2,m_4=1; \mathsf{b}=6) &=
(q= \frac{z (z^4-8 z^2+24)(z^6-18 z^4+144 z^2-480)}{(z^4-12 z^2+48)
(72 z^2-12 z^4+z^6-192)}, \\
p &= \frac{-4 z (z^4-12 z^2+24) (72 z^2-12 z^4+z^6-192)}{
(z^4-8 z^2+24) (-24 z^6+216 z^4-768 z^2+1152+z^8)}) \, ,
\end{split}
\]
which results in $y=(q/z)/(q/z-1)$ equal to
\begin{equation}
y = -\frac{(x^2+4x +6) (-x^3-9 x^2-36 x-60)}{(12  x^3+x^4+54 x^2+96
x+72)}\, ,
\label{y2itemB}
\end{equation}
which also 
follows from equation \eqref{hatw11m1} by flipping the sign
of $x$.
\end{exmp}

We will now discuss other choices for the transformation $M$
and compare  them to results obtained by acting 
with B\"acklund
transformations $\pi,s_3, s_4$ on 
$\alpha_{n;\mathsf{a}}$ from equation 
\eqref{alphapi1}. We will find for $\pi, s_4$ that the resulting parameters 
will agree with those obtained from relations \eqref{Mequality}
with $M_4=\pi^{-1}  s_4, M_{12}= s_1 s_2$, 
respectively, each with two degenerated solutions 
entering inequality  \eqref{Minequality}.
The case of $M_{34}= s_3 s_4$ will be shown  to be equivalent to
$M_1$ although it differs from the sequence obtained by acting with $s_3$.

To trace more easily the effect of these transformations we rename 
the integers 
$n_i \to x_i$, $m_i \to y_i$ for $i=1,2$ to obtain from expression
\eqref{alphapi1}, $2(1+n_2+n_4,-m_2,m_2-n_2,-n_4) $, an expression
\begin{equation}
\pi s_1 \left(\alpha_{n;\mathsf{a}}\right)= \alpha_{m;\mathsf{b}}
=2(1+x_2+x_4,-y_2,y_2-x_2,-x_4)\, ,
\label{alphapi1xy}
\end{equation}
with the consistency condition $x_2 \ge y_2$.

Applying 
$\pi^{-1},s_3, s_4$ on the above relation we get the
following expressions for the B\"acklund transforms
$\alpha_i$ parameters:
\begin{align} 
\pi^{-1}&: 2 (-y_2,y_2-x_2,-x_4,1+x_2+x_4) \,, \label{timespi}\\
s_3&: 2(1+x_2+x_4,-x_2,x_2-y_2,y_2-x_2-x_4) \,,\label{timess3} \\
s_4&: 2(1+x_2,-y_2,y_2-x_2-x_4,x_4)\,. \label{timess4}
\end{align}

Next, we review these expressions in the order they appeared above in
equations \eqref{timespi}-\eqref{timess4} and associate a new B\"acklund
transformations $M_i$ to each of the three cases.  We will be interested in
whether the consistency conditions that will hold for each of the
$M_i$ sequences will be fully derivable from the consistency
condition \eqref{Xoneca} by action of the B\"acklund transformations 
$\pi^{-1},s_3, s_4$ used in the above relations.  If the consistency
relations are mapped into each other together with the parameters
then we will conclude that
the two sequences are fully equivalent and the mapping did not generate
a new degeneracy. 

\subsection{Case of expression \eqref{timespi} with $M_4=\pi^{-1}  s_4$}
\label{subsection:M4}
Perform the following change of variables on variables of equation
\eqref{timespi}:
\begin{equation}
y_2 \to n_2, \; x_4 \to m_4 , \; x_2 \to n_2+n_4-m_4
\label{cov}
\end{equation}
with the condition $x_2 \ge y_2$ transforming into
$n_2+n_4-m_4 >n_2$ or $n_4 \ge m_4$. The condition
$y_4=x_2-y_2$ of $M_1$-sequence is set to consistently transform to $m_2=n_4-m_4$.
This way we obtain :
\begin{equation} 
\alpha = 2(-n_2, m_4-n_4,-m_4,1+n_2+n_4),
\quad \;n_4 \ge m_4 \ge 0 , \;\;n_2, m_4 \in \mathbb{Z}_{+}\, ,
\label{alpha4}
\end{equation}
which is associated with $M_4=\pi^{-1}  s_4$ and
\begin{equation}
\pi^{-1}  s_4 T_2^{-n_2} T_4^{n_4}\vert q=z, p=0 \rangle_{\alpha_{\mathsf{a}}} \ne 
T_2^{-m_2} T_4^{m_4}(q=z,p=0)_{\mathsf{b}} \, ,
\label{x4n}
\end{equation}
with 
\[ \mathsf{a}=2(1+n_4-m_4)=2+2m_2, \;  \mathsf{b}=2(-m_2-n_2)=2 (1-n_2)-
\mathsf{a},
\;\;m_2=n_4-m_4\,.
\]
We see that the model described by $M_1=\pi s_1$ with its condition
$n_2\ge m_2$ is being mapped into a model  described by $M_4=\pi^{-1} s_4$
with $n_4 \ge m_4$ with 
only difference that negative ${\mathsf{a}}$/positive $\mathsf{b}$
transforms into positive ${\mathsf{a}}$/negative $\mathsf{b}$.
Thus with consistency conditions being mapped into each other 
the two sequences are fully equivalent.
This will be illustrated in the following example.
\begin{exmp} 
\label{exmp:x4}
Let us choose 
	\[  m_4=0, \;n_4=1,\; n_2=1,\; \to \;\mathsf{a}=4,\; \mathsf{b}=-4,\;
	m_2 =n_4-m_4=1 \,.
	\]
		
	The corresponding solutions are :
	\begin{equation} \pi^{-1} s_4 T_4^{1} T_2^{-1} 
	\big\vert q=z,\, p=0  \big\rangle_{\alpha_{\mathsf{a}=4}}
	\ne  T_2^{-1} T_4^{0} 
	\big\vert q=z,\, p=0  \big\rangle_{\alpha_{\mathsf{b}=-4}}
	\label{expiis4}
	\end{equation}
	with $\alpha_i=( -2,-2,0, 6)$ holding for both sides.
	
	We find for the left hand side of inequality \eqref{expiis4}:
	\[ 
	q= -\frac{2 z (-4z^2+z^4+8)}{(-2+z^2) (-8z^2+z^4+8)}, \quad
	p= \frac{2 z (-8 z^2+z^4+8)}{(z^2-4)(-4z^2+z^4+8)} \, ,
	\]
	while on the right hand side of  \eqref{expiis4} we find:
	\[ q=z, \; p = \frac{2 z}{-4-z^2} \, ,
	\]
and indeed both solutions satisfy the Painlev\'e V  Hamilton equations
\eqref{hameqs} with $\alpha_i=2 ( -1,-1,0, 3)$.

Corresponding to the above parameters we find by inverting 
relations \eqref{cov} that
$x_2=2 > y_2=1$ and
$x_4=0$. Further, since the condition $m_2=n_4-m_4$ transforms into
$y_4=x_2-y_2$ we get $y_4=1$
for the $M_1=\pi s_1$ sequence.
It follows that the corresponding parameter
found from expression \eqref{Mequality} is $\alpha_i=2(3,-1,-1,0)$.
Next we find that the corresponding solutions of \eqref{Minequality}  
for $M_1=\pi s_1$ sequence are
\[\begin{split} \pi s_1 \mathbb{T} 
\left(x_2=2,x_4=0;\mathsf{a}=-2\right)&= 
\pi s_1 T_2^{-2} \vert q=z, p=0 \rangle_{\alpha_{\mathsf{a}=-2}} \\
&= \vert ( q= \frac{z^6 + 6 z^4}{z (z^4+4 z^2)},
    \quad
    p= z\rangle_{( 6, -2, -2, 0)} 
\, ,
\end{split}
\]
versus
\[ 
\begin{split}
&\mathbb{T} \left(y_2=1,y_4=1; \mathsf{b}=4 \right)=
 T_2^{-1} T_4   \vert q=z, p=0 \rangle_{\alpha_{\mathsf{b}=4}}
 \\&=  \vert 
q= \frac{z \left(z^2-2\right) \left(z^4-8 z^2+24\right)}{\left(z^2-4\right)
\left(z^4-4 z^2+8\right)},
p=  -\frac{2 z \left(z^4-4 z^2+8\right)}{\left(z^2-2\right) \left(z^4-8 z^2+8\right)}
   \rangle_{( 6, -2, -2, 0)} \, , \end{split}
\]
with both solutions of the Painlev\'e V equations \eqref{hameqs}
sharing the same parameters
\begin{equation}
    \alpha_i = \left( 6, -2, -2, 0 \right) \, .
\end{equation}
Thus, as announced, we have been able to map two solutions of $M_1$
and $M_4$ sequences into each other.
\end{exmp}	

\subsection{ Case of expression \eqref{timess3},  $s_3(M_1)$
versus $M_{34}=s_3  s_4$}
\label{subsection:M34}
Here we consider $s_3 (\alpha_i)$ given in the equation  \eqref{timess3}
and we will show that although it agrees with the parameters $\alpha_i$ 
given in formula
\eqref{Mequality} when derived from expression \eqref{Minequality}
with $M_{34}=s_3 s_4$ the consistency conditions will not match.
To study $M_{34}=s_3 s_4$  we consider  the inequality 
\[s_3 s_4 T_2^{-n_2} T_4^{n_4}\,  \vert q=z, p=0 \rangle_{\alpha_{\mathsf{a}}} \ne
 T_2^{-m_2} T_4^{m_4}\, 
 \vert q=z, p=0 \rangle_{\alpha_{\mathsf{b}}} \, .
\]
For parameters of solutions on both sides of this inequality to be
equal we need to have
\begin{equation}\begin{split}
&s_3 s_4 T_2^{-n_2}T_4^{n_4}(\mathsf{a},0,0,2-\mathsf{a})
= s_3 s_4 (\mathsf{a}+2n_2,
-2n_2,-2n_4,2-\mathsf{a}+2 n_4)\\&= (2+2n_2+2
n_4,2-\mathsf{a}-2n_2;\mathsf{a}-2,-2 n_4)\\
&=T_2^{-m_2}T_4^{m_4} (b,0,0,2-b)=
(\mathsf{b}+2m_2,-2m_2,-2m_4,2-\mathsf{b}+2m_4)\, .
\end{split}
\label{s3s4equality}
\end{equation}
Solving for $\mathsf{a}$ and $\mathsf{b}$ yields
\begin{equation} \mathsf{a}=2-2m_4=2+2m_2-2n_2,\; \; \mathsf{b}=2+
2m_4+2n_4=4+2n_2 - \mathsf{a} >0 \, ,
\label{aM34}
\end{equation}
with the consistency relation 
\begin{equation} m_4=n_2-m_2\,,
\label{s3s4consistency}
\end{equation}
required for the above equations to hold.

We notice that this consistency relation 
ensures that $\mathsf{b}$ is always positive.

Inserting the values of  $\mathsf{a}$ and $\mathsf{b}$ back into 
the relation \eqref{s3s4equality} we obtain:
\begin{equation}
\alpha_i= 2 (1+n_2+n_4,-m_2,m_2-n_2,-n_4)\, ,
\label{s3s4alpha}
\end{equation}
in full agreement with equation  \eqref{timess3} reproduced below:
\[
s_3(\alpha_i) = 2(1+x_2+x_4,-x_2,x_2-y_2,y_2-x_2-x_4)\,, 
\]
when we identify $x_2=m_2$, $y_2=n_2$ , $x_4= n_2+n_4-m_2$. Note however
that since $n_2-m_2\ge 0$ it follows that \eqref{s3s4consistency}
reads in terms of these variables as: $y_2-x_2 \ge  0$, which is
just opposite to the original condition $x_2 -y_2 \ge 0 $ of the 
$M_1$-sequence seen below
\eqref{timess3}. Thus this time the consistency relations did not get
mapped into each other. 

Does this result mean that the $M_{34}$-sequence is independent of the
$M_1$-sequence because $s_3$ failed to connect those two
cases? It turns out that $M_{34}$-sequence is fully equivalent to 
$M_1$-sequence because of relation $s_3 s_4 =\pi s_1 T_1^{-1}$, which
is a special case of relations \eqref{nextsi}. It follows from this
relation that
\begin{equation}
\begin{split}
s_3 s_4 T_2^{-n_2} T_4^{n_4}\,  
\vert q=z, p=0 \rangle_{\alpha_{\mathsf{a}=2-2m_4}} &=
\pi s_1 T_1^{-1}T_2^{-n_2} T_4^{n_4}\,  
\vert q=z, p=0 \rangle_{\alpha_{\mathsf{a}=2-2m_4}}\\
&=
\pi s_1 T_2^{-n_2} T_4^{n_4}\,  
\vert q=z, p=0 \rangle_{\alpha_{\mathsf{a}=-2m_4}} \, ,
\end{split}
\label{M34=M1}
\end{equation} 
where we inserted the value of $\mathsf{a}$ from relation \eqref{aM34}
and used relation \eqref{T1shift}.
The above expression is equal to the one given in equation \eqref{Xone}
then one takes into account the value of the parameter $\mathsf{a}$
given in \eqref{Xonec}. Thus the $M_{34}$-sequence is fully equivalent to
the $M_1$-sequence.

It is still warranted to consider the sequence generated by action of
$s_3$ on the $M_1$-sequence. The following observation is crucial. Consider
$\alpha_i=(\alpha_1,\alpha_2,\alpha_3,\alpha_4)$ entering expressions
for the parameters
$\alpha = \alpha_3^2/8$, $\beta=-\alpha_1^2/8$ and
$\gamma = (\alpha_2-\alpha_4)/2$ of the Painlev\'e V equation \eqref{finalPV}.
The B\"acklund transformation $s_3$ transforms $\alpha_i$ into
$(\alpha_1,\alpha_2+\alpha_3,-\alpha_3,\alpha_4+\alpha_3)$ 
maintaining the parameters
$\alpha, \beta, \gamma$  of the Painlev\'e V equation \eqref{finalPV}
clearly invariant.
Note that the remaining B\"acklund transformations $s_1,s_2,s_4$ will
all change the parameters
$\alpha , \beta, \gamma$. 
However the $s_3$ transforms $q,p$ as follows 
\[s_3: q \to q, \; p \to p -\frac{\alpha_3}{z-q}, 
\]
and accordingly will leave the solution $y$
of the Painlev\'e V
equation \eqref{finalPV} invariant.
To illustrate these considerations we will act with
$s_3$ on configurations given in example \ref{exmp:n2=2n4=2}.
\begin{exmp}
\label{exmp:n2=1n4=3}
As an example we consider acting  with $s_3$ on \eqref{solutions2}, 
which transforms parameters
as follows:  $2(5,-1,-1,-2) \to 2(5,-2,1,-3)$
Accordingly, we deal with the case of  
\begin{equation} 
 n_2=1,\, n_4=3, \; m_2=2 \; \to m_4=n_2-m_2=-1,
 \, \alpha_i=(5,-2,1,-3).
\label{itemC}
\end{equation}
We note that now  $m_4=n_2-m_2$ is negative, however
the corresponding coefficients of the Painlev\'e V equation, 
for $\epsilon=1$, are the ones in \eqref{paramPVbar1}
as seen in Example \ref{exmp:n2=2n4=2}.
Acting  with $s_3$ on solution 
\eqref{solutions11} we get
\begin{equation}
q= \frac{z (z^2-6) (z^4-16 z^2+80)}{(z^2-8)(z^4-12 z^2+48)}, \;\;
 p =  \frac{(z^4-12 z^2+48)}{ z(z^2-6)} \, ,
\label{solutions111}
\end{equation}
while acting  with $s_3$ on \eqref{solutions2}, 
we get
\begin{equation}
\begin{split}
q &=
z\,\frac{(8\,z^2+24+z^4)\,(z^6+18\,z^4+144\,z^2+480)}{(z^4+12\,z^2+48)\,(z^6+12\,z^4+72\,z^2+192)}
\, ,\\
p &= -4\,\frac{(z^4+12\,z^2+48)}{(z\,(8\,z^2+24+z^4)} \, .
\label{solutions1}
\end{split}
\end{equation}
Solutions \eqref{solutions111} and  \eqref{solutions1} satisfy the Painlev\'e V
Hamilton equations \eqref{hameqs} 
with $\alpha_i=(10,-4,2,-6)$ that differ from solutions in Example 
\ref{exmp:n2=2n4=2}, which satisfy the Painlev\'e V Hamilton equations 
with the $\alpha_i=2(5,-1,-1,-2) $. 
However, since $s_3 (10,-4,2,-6)=2 (5,-1,-1,-2) $ and $s_3(y(x))=y(x)$, 
they give rise to the identical  solutions $y(x)$ as obtained
in Example 
\ref{exmp:n2=2n4=2} for the Painlev\'e V equation \eqref{finalPV} 
with the coefficients \eqref{paramPVbar1}.
\end{exmp}
\subsection{ Case of expression \eqref{timess4} with $M_{12}= s_1s_2$}
\label{subsection:M12}
In this case 
we consider $s_4 (\alpha)$ from equation
\eqref{timess4} and compare with an expression for the 
$\alpha$ that we obtain from \eqref{Mequality} for $M=M_{12}$:
\begin{equation} \begin{split}
\alpha &= T_2^{-m_2} T_4^{m_4}
(\mathsf{b},0,0,2-\mathsf{b})
=T_2^{-m_2} T_4^{m_4}\,\left(\mathsf{b},0,0,2-\mathsf{b}\right) \\
&= \left( \mathsf{b}+2 m_2, -2 m_2, -2 m_4,
2- \mathsf{b}+2 m_4 \right) \\
&=s_1s_2 T_2^{-n_2} T_4^{n_4}
(\mathsf{a},0,0,2-\mathsf{a})= \left( -\mathsf{a},\mathsf{a} +2 n_2,
-2 n_2 -2 m_2, 2+2 n_4 \right) \, .
\label{alpha12}
\end{split}
\end{equation}
The consistency requires this time that:
\begin{equation}
m_4 =n_2+n_4 \, ,
\label{consistencyM12}
\end{equation}
which leads to the following expressions:
\[ \mathsf{a}=-2 n_2-2m_2,\; \mathsf{b}=2 n_2 = -2m_2 -\mathsf{a}\, . 
\]
Plugging these values back into equation \eqref{alpha12} we
obtain an expression for $\alpha$:
\begin{equation}
\alpha=  2(n_2+m_2,-m_2,-n_2-n_4,1+n_4) \;
\quad \; m_2, n_4 \in \mathbb{Z}_{+}, \; 
\label{alpha12a}
\end{equation}
that also follows from inequality \eqref{Minequality} with $M_{12}=s_1 s_2$:
\begin{equation}
s_1s_2 T_2^{-n_2} T_4^{n_4}\Big\vert q=z, p=0 \Big\rangle_{\alpha_{\mathsf{a}}} \ne
T_2^{-m_2} T_4^{m_4}\Big\vert q=z, p=0 \Big\rangle_{\alpha_{\mathsf{b}}}, \;m_4=n_2+n_4 \,.
\label{Ms1s2}
\end{equation}
Expression \eqref{alpha12} agrees with the result of \eqref{timess4} 
for :
\[
m_2= y_2,\; n_4=x_4-1, \; n_2= x_2-y_2+1
\]
Thus the coefficients $x_2, x_4, y_2$ need to satisfy inequalities
$x_4 \ge 1, x_2 \ge y_2$, which are consistent with 
conditions \eqref{Xonec}.
Note that $x_2+ 1 > y_2$ always holds  since $x_2 \ge y_2$ and
accordingly $n_2 >0$.

We see that both sequences will map into each other when $x_4$
variable of the $M_1$ sequence takes values $x_4=1,2,{\ldots} $ and 
correspondingly the
$n_2$ variable of the $M_{12}$ sequence takes values $n_2=1,2,{\ldots} $.

\section{Discussion}
\label{section:discussion}
We have examined the cases of two-fold degeneracy of the Painlev\'e V
rational solutions 
connected with the B\"acklund
transformations $M_1=\pi s_1, M_4=\pi^{-1} s_4, M_{34}= s_3 s_4,
M_{12}=s_1 s_2$ that enter the basic inequality \eqref{Minequality}
that relates the two degenerated solutions with the equal parameter 
\eqref{Mequality} and showed that all four sequences of degenerated
solutions are fully equivalent by employing B\"acklund transformations 
$\pi^{-1}$ and $s_4$ to show equivalence of $M_1$-sequence with those
of $M_4=\pi^{-1} s_4, M_{12}=s_1 s_2$ and relation 
$s_3 s_4 =\pi s_1 T_1^{-1}$
for equivalence between $M_1=\pi s_1$ and  $M_{34}= s_3 s_4$.

In number of Examples \ref{exmp:n2=2n4=2}, \ref{exmp:n2=3n4=1}
and \ref{exmp:n2=1n4=3} we have considered solutions
with   the Painlev\'e V coefficients:
\begin{equation}
\alpha = \frac12,\; \; \beta= - \frac{25}{2},\;\; 
\gamma= \pm 1 \, .
\label{4.11coefs}
\end{equation}
Let us now summarize the results of these considerations in the
setting of $M_1$-sequence. 

Recalling the expression \eqref{paramPVbar}  for
the Painlev\'e V  equation coefficients with $\epsilon=1$ 
 and inserting the relevant 
components of $\alpha_i$ \eqref{alphapi1} into these expressions 
we find that in order to match them with the expression \eqref {4.11coefs}
we need to solve the following three equations 
\begin{equation} 
(1+n_2+n_4)^2=25, \; (m_2-n_2)^2=1, \; (n_4-m_2)^2=1\, ,
\label{3eqs}
\end{equation}
for the three variables $n_2,n_4,m_2$ that all need to be positive integers.

Equations \eqref{3eqs} have $8$  solutions in total but only half of
them with positive integers $n_2,n_4,m_2 \in \mathbb{Z}_{+}$. 
We list these $4$ relevant 
solutions below:
\begin{enumerate}
\item[A)] $n_2=n_4=2$, \; $m_2=1$ \; $\to$ $m_4=n_2-m_2=1$,
\,$\gamma=1$, $\, \alpha_i=2(5,-1,-1,-2)$.
\item[B)] $n_2=3, \, n_4=1$, \; $m_2=2$ \; $\to$ $m_4=n_2-m_2=1$,
\,$\gamma=-1$, $\, \alpha_i=2(5,-2,-1,-1)$.
\item[C)] $n_2=1,\, n_4=3$, \; $m_2=2$ \; $\to$ $m_4=n_2-m_2=-1$,
\,$\gamma=1$, $\, \alpha_i=2(5,-2,1,-3)$.
\item[D)] $n_2=2, \, n_4=2$, \; $m_2=3$ \; $\to$ $m_4=n_2-m_2=-1$,
\,$\gamma=-1$, $\, \alpha_i=2(5,-3,1,-2)$.
\end{enumerate}
Items $A)$  and $B)$ have been discussed in
Examples \ref{exmp:n2=2n4=2} and \ref{exmp:n2=3n4=1}, where we noticed
that they satisfy the condition $n_2 \ge m_2$ (or $m_4
\ge 0$) and are  therefore a part of the $M_1$-sequence. 

We have seen that on the level
of Painlev\'e V equation \eqref{finalPV} the transformation of 
solutions obtained inside the $M_1$-sequence 
with the  parameters listed in case A) 
to solutions of case B)
was fully accomplished by flipping $\gamma \to -\gamma$ or equivalently 
flipping $x \to -x$. On the level of the Hamilton Painlev\'e V
equations the corresponding $q,p$ solutions 
solve the equations \eqref{hameqs} with different $\alpha_i$
given above in A) and B).  Recall that in \cite{AGLZ} we have introduced $x$
as $x=z^2/(2 \epsilon)$ with $\epsilon^2=1$. Thus here we are exercising the
freedom of changing a sign of $\epsilon$ that changes a sign of
$\gamma$ (see again  \cite{AGLZ}).

The cases C) and D) are mapped from 
A) and B) by action of $s_3$:
\[
C) = s_3(A)), \quad 
D) = s_3(B)), \]
as can be verified by inspecting the parameters $\alpha_i$.
We have seen the case C) being discussed in Example \ref{exmp:n2=1n4=3}. 
Each of these two cases exhibits therefore the two-fold degeneracy of the Hamilton
Painlev\'e V equations with solutions that are an $s_3$ image of the 
corresponding solutions of $M_1$-sequence with parameters 
of case A) and B).  
Since  $s_3$ keeps both the 
coefficients and the solution of the Painlev\'e V equation
\eqref{finalPV} invariant, we conclude that 
the  Painlev\'e V solutions associated
to cases C) and D) are fully equal to those already  found in cases 
A) and B).

In all examples we have seen $\mathsf{a}$ and $\mathsf{b}$ are even
integers and having (to some degree) an opposite sign.
For the $M_1$-sequence $\mathsf{a}  \le 0$ and $\mathsf{b} \ge 2$
and such that $\mathsf{a}/2+\mathsf{b}/2=1,2,{\ldots} $.
For the $M_{12}$ sequence $\mathsf{a}  \le 0$ and $\mathsf{b} \ge 0$
and $\mathsf{a}/2+\mathsf{b}/2=0,-1,-2,{\ldots} $.
For the $M_4$ sequence $\mathsf{a}  \ge 0$ and $\mathsf{b} \le 0$
such that $\mathsf{a}/2+\mathsf{b}/2=0, -1,-2,{\ldots} $.
For the $M_{34}$-sequence it holds that $\mathsf{a}  \le 2$ and $\mathsf{b} \ge 2$
and $\mathsf{a}/2+\mathsf{b}/2=2,3,{\ldots} $ as expected since
the $M_{34}$-sequence is equivalent to the $M_1$-sequence only with
$\mathsf{a}$ shifted by $2$.

As we have noted in section \ref{section:background} the value of the
parameter $\mathsf{a} $ can be shifted by an even integer $2n$ through
the action of $T_1^{n}$. For degenerated solutions one can use this
freedom to set, for example, the parameter  $\mathsf{a} $ 
to zero since it is an even integer. However the same operation will raise
or lower the value of the connected parameter $\mathsf{b} $ and therefore 
maintain  invariant the value of their
sum.

\begin{exmp}
\label{example:discussion}
As an example consider $\mathsf{a}$ and $\mathsf{b}$ such that $\mathsf{a}=-2n$
and $\mathsf{b}=2n+2 k$ for $n\in \mathbb{Z}$ and $k=1,2,3,{\ldots} $.
Comparing with the paragraph above we see that this case fits into the
$M_1$-sequence of degenerated solutions. Comparing with the expressions
\eqref{Xoneca} and \eqref{Xonec} we find that $n_4=k-1$ and $m_4=n$.
We conclude that for any fixed integers $n \ge 0$ and $k>0$ we find a pair of solutions
belonging to $M_1$-sequence:
\[
\pi s_1 \mathbb{T} (n_2,k-1;\mathsf{a}=-2n) \quad {\rm and} \quad
\mathbb{T} (n_2-n,n;\mathsf{b}=2n+2k)\, ,
\] 
that satisfy the Painlev\'e V equations with the same parameters
\begin{equation}
\alpha_i= 2( 1+n_2+n_4,-m_2,m_2-n_2,-n_4)=
2(n_2+k,n-n_2,-n, 1-k)\, ,
\label{degealpha}
\end{equation}
valid for any integer $n_2$ such that $n_2 \ge n$.
\end{exmp}
Comparing $\alpha (m; \mathsf{b})=(\mathsf{b} +2 m_2,\,
-2 m_2,\,
  -2 m_4,\,  2-\mathsf{b} +2 m_4)$ from expression 
\eqref{orbit}.
we recognize that it agrees with expression for the parameter 
\eqref{degealpha} for $\mathsf{b}= 2(k+n)$ and $m_2=n_2-n\ge 0$,
$n=m_4$. 

In summary, we have developed an explicit construction that applies
to the two-fold degeneracy of Painlev\'e V Hamilton equations  and determines the two
degenerated solutions and the parameters  of Painlev\'e V equations that
they share. We also found a condition
for a solution $\mathbb{T} (m; \mathsf{b})$ on the orbit of
$T_2^{-m_2}T_4^{m_4}$ to agree with one of the two degenerated solutions
and the condition is that the parameter 
$\mathsf{b}$ is an even integer ( a positive integer  
for the $M_1$-sequence and a negative for  the $M_4$-sequence).

Recall that the Painlev\'e V Hamilton system is closely related to the dressing
chain of even, $N=4$ periodicity, see \cite{AGLZ} and references therein.
Our discussion based on translation operators 
indicates that degeneracy will exist for 
all dressing chains of even periodicity because of existence of 
exclusion rules for translation operators permitted to act
on special types of seed solutions. Especially,
it will occur for  $N=6$ periodic dressing chain discussed in \cite{AGLZ}.
A natural problem to investigate
is whether a degree of degeneracy
(how many solutions will share the parameter $\alpha_i$)
will change in case of higher dressing chains of even period $N>4$.

\subsection*{Acknowledgements}

This study was financed in part  by the Coordena\c{c}\~{a}o de
Aperfei\c{c}oamento de Pessoal de N\'ivel Superior - Brasil (CAPES) - Finance Code 001
(G.V.L.) and by CNPq and FAPESP (J.F.G. and A.H.Z.).

\label{lastpage}
\end{document}